\documentclass[%
 aip,
 jmp,%
 amsmath,amssymb,
 reprint,%
]{revtex4-1}

\usepackage{graphicx}
\usepackage{dcolumn}
\usepackage{bm}

\begin{document}

\preprint{AIP/123-QED}

\title {About relationship between flux and concentration gradient of particles 
at description of diffusion with the usage of random walk model}

\author{M.N. Ovchinnikov}
 \affiliation {Physics Department, Kazan Federal University.}
 \email{Marat.Ovchinnikov@kpfu.ru}

\date{\today}

\begin{abstract}
The fundamental solutions of diffusion equation for the local-equilibrium and nonlocal models are considered as the limiting cases of the solution of a problem related to consideration of the Brownian particles random walks. The differences between fundamental solutions were studied. It was shown that on the period of observation time and distances exceeding the time and space associated with one step of random walk the fundamental solutions of diffusion and telegraph equations are very close to each other. In particular, these fundamental solutions are very close to the values of the probability density distribution of diffusive particles that is obtained from the accurate solution of the random walk problem. The difference between the probability density values for the models considered is decreased with time as $t^{-3/2}$, the relative difference as $t^{-1}$, and velocity of propagation of these fixed distribution densities is proportional to $t^{1/2}$. The new modified non-local diffusion equation is suggested. It contains only microparameters of the random walk problem. 

\end{abstract}

\maketitle

\section{\label{sec:level1}Introduction }

It is known that one of the properties of linear diffusion equation solution 

\begin{eqnarray}
 \frac{{\partial C(\vec x,t)}}{{\partial t}} = D\Delta C(\vec x,t)
\label{eq:1}
\end{eqnarray}
is the infinite velocity of perturbations propagation. It is seen, for example, from the fundamental solution of this equation 

\begin{equation}
 \rho _G (\vec x,t) = \frac{{\Theta (t)}}{{\sqrt {4\pi Dt} }}\exp ( - \frac{{\left| {\vec x} \right|^2 }}{{4Dt}}).
\end{equation}
Here  $ \rho_G (\vec x,t)\ $ is the fundamental solution presented in the Gaussian form, $C$ is concentration, $D$ is the diffusion coefficient, $\Theta$ defines the unit step Heaviside function, $x$ is associated with current coordinate, $t$ is the current time. This solution contradicts to intuitive speculations that diffusive particles and any information about their movement are propagated with finite velocities. We should mark that equation ~(\ref{eq:1}) is a consequence of supposition that principles associated with locality and local thermodynamic equilibrium are valid. In the frame of this supposition one can use the diffusion Fick's law in standard form 

\begin{equation}
 \vec J(\vec x,t) =  - D\nabla C(\vec x,t)
\end{equation}
connecting the current particle function and the concentration gradient. By analogy, for conductive heat transfer one can write the Fourier law connecting the heat flux with temperature gradient and for liquids and gases involved in the filtration phenomenon in porous medium the Darcy's law connecting the stream of liquid with pressure gradient. 

One of the approaches related to resolution of the problem connected to corrections for the infinite velocity of perturbations propagation lies in the usage of linear nonlocal models, for example, suggested by the extended of nonequilibrium thermodynamics \cite{Jou1999},
\cite{Sobol1997}. In this case in expression (3) some additional relaxation terms are introduced. In the simplest case, the Fick's law in the nonlocal interpretation is written as

\begin{equation}
\vec J(\vec x,t) + \tau \frac{{\partial \vec J(\vec x,t)}}{{\partial {\kern 1pt} t}} =  - D\nabla C(\vec x,t)
\end{equation}
and equation ~(\ref{eq:1}) is reduced to the telegraph equation of the type

\begin{equation}
\frac{{\partial (\vec x,t)}}{{\partial t}} + \tau \frac{{{\partial ^2}(\vec x,t)}}{{\partial t{}^2}} = D\Delta (\vec x,t).
\end{equation}

\noindent The fundamental solution of the last one for $1D$ case is expressed as 

\begin{eqnarray}
{\rho _{TE}}(x,t) = \frac{1}{{\sqrt {4D\tau } }}\exp ( - \frac{t}{{2\tau }})\,\Theta \left( {\sqrt {\frac{D}{\tau }} t - x} \right) \nonumber\\
{I_0} \left( {\frac{1}{{2\tau }}\sqrt {{t^2} - \frac{{{x^2}\tau }}{D}} } \right)\
\end{eqnarray}
where $I_0$ is the modified Bessel's function of the first kind having zeroth order. The paradox of the infinite velocity is eliminated and the limiting velocity describing the perturbations propagation in a nonequilibrium medium becomes finite and is equaled to $ V = \sqrt {{D \mathord{\left/ {\vphantom {D \tau }} \right. \kern-\nulldelimiterspace} \tau }} $.

In the given paper the proper selection of the fundamental solutions plays a key role. Their knowledge determines the specific solutions of linear equations of mathematical physics as well, because the corresponding Green's functions can be expressed in the form of linear combinations of fundamental solutions including their derivatives and integrals. 

Equations of type ~(\ref{eq:1}) for many real situations of macroscopic physics are good approximation. But together with this statement the consideration of the nonlocal effects is an actual problem for relatively small intervals of observation times and distances. It is interesting to consider the evaluation of the limits of applicability of the local thermodynamic equilibrium hypothesis and the finding of approximate approaches related to determination of the current of the value studied and gradient of potential of the corresponding scalar field. Taking into account the fact that approach (4) is also approximate, it would be interesting to evaluate errors which arise in the usage of similar expressions in comparison with solution of the "exact" transfer equation which is remained unknown. 

The certain complexity in experimental testing in laboratory conditions the adequacy of equations ~(\ref{eq:1}) and (4) lies in nonlinear dependencies of the key parameters involved. It creates some difficulties in interpretation of the results obtained. So, in the given paper the dynamics of hypothetical linear systems modelling the transfer phenomenon is considered. 

It is necessary to remind here about the inertia heat phenomenon when nonlinearity leads also to the finite velocities of perturbations \cite{Kurd1995} but similar and relatively exotic models are out of the scope of this research. 

If one come back to micro-level the diffusion phenomenon, conductive heat transfer and filtration of liquids and gases in porous media and similar phenomena can be described with the help of random behavior of microparticles. For diffusion and heat transfer phenomenon one can introduce concepts as the characteristic length $\Delta x$ and time $\Delta t$ of free particles path or quasiparticles of phonon type; for filtration we consider the characteristic distances and times associated with the hopping process of liquid droplets from one pore to another one. 

Hence, from microscopic point of view of the phenomena considered it is important to establish the correlations between phenomenological parameters of the corresponding theories similar to the extended irreversible thermodynamics with characteristic relaxation times τ and microparameters of the transfer process studied. 

In random walk model the hopping length  $\Delta x$ and hopping time $\Delta t$ characterize the dynamics of the system considered on microlevel. For various media and processes these microparameters are varied in wide limits. The length of free path for phonons in crystals is changed in the range $10^{-7}-10^{-10}$ m that corresponds to the free path time located in the range $10^{-10}-10^{-13}$ s for frequencies of the phonon spectrum from the interval $10^{10}-10^{12}$ Hz. We should note that in paper \cite{Volz1996} with the usage of the molecular dynamics methods it was shown that characteristic relaxation times in nonlocal heat transfer model written in the form of telegraph equation (5) for solid argon crystal are very small ($10^{-11}$ s) that comparable with times of the phonons free path in a crystal. The characteristic microscopic hopping length in diffusion phenomenon in the most cases is located in the range $10^{-3}-10^{-9}$ m, for filtration – $10^{-7}-10^{-3}$ m and the corresponding characteristic times are located in the intervals $10^{-10}-10^{0}$ s  and $10^{-3}-10^{-1}$ s, accordingly.

\section{\label{sec:level1}The model}

The random walk model at present time is widely used for description of many processes; in particular, the fractal dynamic models are developed in \cite{Met2000}. Supposing that on validity not only nice equations can pretend but also the nice algorithms, as well, we will be based on supposition that true description of diffusion phenomenon is determined by random walk model and there is unknown linear equation which correctly describes of the transfer phenomenon of diffusion type. 

In this paper the simplest variant of the random walk model for $1D$ system is considered. In this model the hoppings of a particle (or quasiparticles) are realized with equal probabilities during the interval of time $\Delta t$  for distances $+\Delta x$  or  $-\Delta x$ . Then for the discrete variant of the problem formulated one can obtain the solution for the probability $P$ of a particle location in coordinate $x_m$  in the moment $t_n$  in the form \cite{Haken1978}

\begin{equation}
P({x_m},{t_n}) = \frac{{({t_n}/\Delta t)!}}{{\left( {\frac{{({t_n}/\Delta t + {x_m}/\Delta x)}}{2}} \right){\kern 1pt} {\kern 1pt} {\kern 1pt} !\left( {\frac{{({t_n}/\Delta t - {x_m}/\Delta x)}}{2}} \right)\,!\;{2^{\left( {{t_n}/\Delta t} \right)}}}}
\end{equation}
It is known that expression (2) presents one of the partial cases of solution of the random walk problem describing the probability density distribution of a random value. Really, if divide (7) on $2\Delta x$ and put

\begin{equation}
\frac{{{{\left( {\Delta \,x} \right)}^2}}}{{2\Delta \,t}} \to D = const > 0
\end{equation}
at $\Delta x\to 0, \Delta t\to 0$, then one can obtain expression for the probability density function describing a probability of location of diffusive particle in the vicinity of the point \cite{Haken1978} $x$ in the form (2). In this case, simple calculations show that the velocity $V$ of perturbations propagation accepts the infinite value ($V = {\text{lim}}(\Delta \,x/\Delta t) \to \infty $ at $\Delta x\to 0, \Delta t\to 0$  ) but the solution itself has automodel character with respect to variable ${x^2}/t$. We should note that in \cite{Haken1978} for simplification of expression (7) containing factorials with the usage of the Stirling's decomposition formula 

\begin{eqnarray}
  z! \approx {z^z}\exp ( - z)\sqrt {2\pi \,z}  
  \left\{ {1 + \frac{1}{{12z}} + \frac{1}{{288{z^2}}} + 0({z^{ - 3}})} \right\} 
\end{eqnarray}
only the first term in the braces is kept. It allows to introduce a criterion for correct evaluation of this approach in the form of the following inequalities 

\begin{subequations}
\label{eq:whole}
\begin{equation}
t_/ \Delta t >> 1, \label{subeq:1}
\end{equation}
\begin{equation}
x_m << t_n \Delta x / \Delta t . \label{subeq:2}
\end{equation}
\end{subequations}

The fundamental solution for telegraph type equation (4) represents another limiting case of the random walk problem \cite{Kac1974}. In this case  ${\text{lim}}(\Delta \,x/\Delta t) =V < \infty $ at $\Delta x\to 0, \Delta t\to 0$.   
The selection of fundamental solutions as a ground for analysis of transfer processes is justified by the fact that for linear equations of mathematical physics the knowledge of a fundamental solution determines specific solutions of these equations as well, if the corresponding initial and boundary conditions are known. It is related to the fact that the corresponding Green functions can be expressed in the form of linear combinations of fundamental solutions including their derivatives and integrals. 
Then at transition to continuous variables it is convenient to measure a current length in the unit of elementary length $\Delta x$ and the current time in the units of $\Delta t$. We put also $V=\Delta x/\Delta t=1$, $D=1/2$, $\tau=V^{2}/D= \Delta t/2 = 1/2$.

Then the fundamental solution of equation ~(\ref{eq:1}) is written as 
\begin{equation}
{\rho _G}(x,t) = \frac{1}{{\sqrt {2\pi t} }}\exp ( - \frac{{{x^2}}}{{2t}})
\end{equation}

and fundamental solution for (4) as

\begin{equation}
{\rho _{TE}}(x,t) = \exp ( - t)\,{I_0}\left( {\sqrt {{t^2} - {x^2}} } \right)\,\Theta {\kern 1pt} {\kern 1pt} (t - \left| x \right|)
\end{equation}
and the probability density distribution function of a particle located in the vicinity of the coordinate $x$ and in the moment time $t$ for model (5) and considered in the given paper as the fundamental solution of unknown transfer equations is written as

\begin{equation}
{\rho _{RW}}(x,t) = \frac{{(t)!}}{{2\;\left( {\frac{{(t - x)}}{2}} \right){\kern 1pt} {\kern 1pt} {\kern 1pt} !\left( {\frac{{(t + x)}}{2}} \right)\,!\;{2^t}}}
\end{equation}
Or, being presented in the form of the Euler's gamma-functions it can be presented as 

\begin{equation}
{\rho _{RW}}(x,t) = \frac{{\Gamma (t + 1)}}{{\Gamma \;\left( {\frac{{(t - x)}}{2} + 1} \right)\,\Gamma \,\left( {\frac{{(t + x)}}{2} + 1} \right)\;{2^{t + 1}}}}
\end{equation}
Here and below the symbol $G$ (Gauss) indicates on equation Eq.~(\ref{eq:1}), the symbol $TE$ (Telegrapher equation) – on equation  (5), and symbol $RW$ (Random Walk) – on initial equation (7).
Then we will use the following hypothesis: the true solution of evolution equation of the diffusion type presents the random walk solution in the form of fundamental solution (14) but presentations (2, 11) and (6, 12) – present approximate approaches. Solution (14) corresponds to the unknown linear evolution equation.

\section{\label{sec:level1}Results}

Let us consider the evolution of equations (11, 12 and 14). On Figure1 we demonstrate the values of the function $\rho _{RW}$
  end its gradient $\nabla {\rho _{RW}}$
  for the fixed times t= 30 and  100. On Figure2 we show the differences of the function  $\rho _{RW}$ (14) from the functions $\rho _G$
  (11) and  $\rho _{TE}$ (12) calculated at the same values of time.

\begin{figure}
\includegraphics{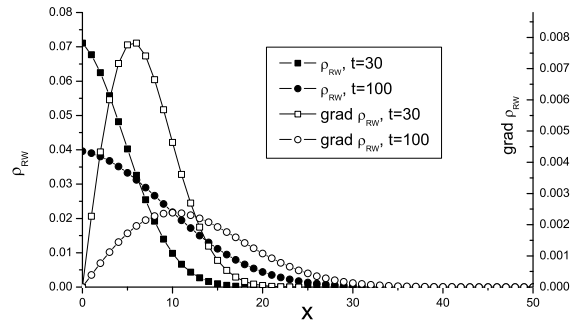}
\caption{\label{FIG1:epsart} Comparison of the values   and    for the fixed values of time t=30 and t=100.}
\end{figure}

\begin{figure}
\includegraphics{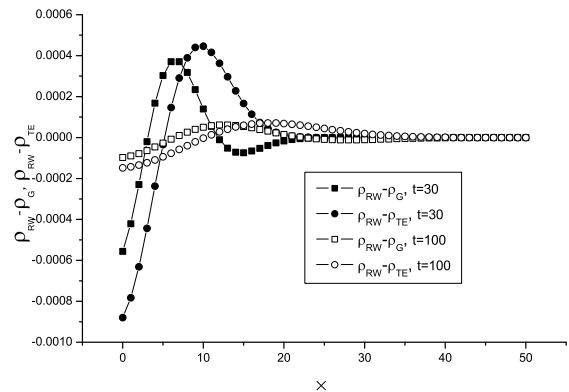}
\caption{\label{FIG2:epsart} The differences of solutions (11), (12) and (14) for the values t=30 and t=100.}
\end{figure}

With increasing of the time the functions considered not changing their forms are decreased in their values and are stretched in the space. The differences between the functions $\rho _G$, $\rho _{TE}$ , $\rho _{RW}$  become small and are decreased with the time growing. 

Taking into account the influence of the second decomposition term figuring in the Stirling's decomposition expression (11) for factorials and the asymptotic behavior of the Bessel function  at ($z>>1$)

\begin{equation}
{I_0}(z) \approx \frac{{\exp (z)}}{{\sqrt {2\pi {\kern 1pt} z} }}(1 + \frac{1}{{8{\kern 1pt} z}} + ...),
\end{equation}
it is easy to calculate the values of the functions considered at the point $x=0$ in different values of time at $t>>1$ . These values are equaled approximately 

\begin{subequations}
\label{eq:whole}
\begin{equation}
{\rho _{RW}}(0,t) \approx \frac{1}{{\sqrt {2\pi {\kern 1pt} t} }}(1 - \frac{1}{{4{\kern 1pt} t}} + ...), \label{subeq:1}
\end{equation}
\begin{equation}
{\rho _{TE}}(0,t) \approx \frac{1}{{\sqrt {2\pi {\kern 1pt} t} }}(1 + \frac{1}{{8{\kern 1pt} t}} + ...),\label{subeq:2}
\end{equation}
\begin{equation}
{\rho _G}(0,t) = \frac{1}{{\sqrt {2\pi {\kern 1pt} t} }} \label{subeq:3}
\end{equation}
\end{subequations}
From these relationships it follows that the differences of the functions (16) are decreased with time as $t^{-3/2}$, their relative differences as $t^{-1}$  and their absolute values as $t^{-1/2}$. The corresponding calculations show that for other values of the current coordinate $x$ the differences of the functions (11, 12 and 14) values are decreased with time as  $t^{-3/2}$. Incidentally, the spatial stretching of the functions themselves at $t >> 1$ is proportional to  $t^{1/2}$. 

Let us consider the gradient concentrations dynamics corresponding to solutions (11), (12) and (14). For the function  $\rho _{RW}$  in notations used the corresponding gradient will look as 

\begin{equation}
\nabla {\rho _{RW}} = \frac{1}{2}\frac{{\Gamma (t - 1)(\Psi ((t - x)/2) - \Psi ((t + x)/2))}}{{{2^{t + 1}}\Gamma ((t - x)/2)\Gamma ((t + x)/2)}}
\end{equation}
where $ \psi (x) = \frac{d}{{dx}}\ln (\Gamma (x))\ $, and for the functions  $\rho _G $, $\rho _{TE}$,  as   

\begin{equation}
\nabla {\rho _G} =  - \frac{x}{t}\frac{1}{{\sqrt {2\pi t} }}\exp ( - \frac{{{x^2}}}{{2t}})
\end{equation}
and

\begin{equation}
\nabla {\rho _{TE}} =  - x\exp ( - t)\,{I_1}\left( {\sqrt {{t^2} - {x^2}} } \right)\,/\left( {\sqrt {{t^2} - {x^2}} } \right).
\end{equation}
For all functions considered the gradient concentrations distributions at the fixed moments of time coincide with the functions that tend to zero at $x\to \infty$  and $x \to 0$  and having one maximum. For equation (18) at the fixed value of time the maximum value should be observed at $x_m = \sqrt {2{\kern 1pt} D{\kern 1pt} t} $ and the value of function (18) in this point is $\nabla \rho {}_G^{}({x_m}) = 1/(\sqrt {8\pi {\kern 1pt} e} {\kern 1pt} D{\kern 1pt} t)$ , the velocity of the movement of the maximum point ${V^G}_m$ and its halfwidth $\Delta ^G_m$ will be equaled to  ${V^G}_m({x_m}) = \sqrt {D/2{\kern 1pt} t} \sim{t^{1/2}}$ and $\Delta ^G_m \approx 1.6\sqrt {2{\kern 1pt} D{\kern 1pt} t} \sim{t^{1/2}}$ , correspondingly. The calculations show that the differences between maximum locations points of functions ${\rho _G}$ , ${\rho _{TE}}$  и ${\rho _{RW}}$  and their relative differences are decayed as ${t^{ - 1}}$  and the relative changes of the maximum values of the corresponding gradients  $\nabla {\rho _G}(x,t)$, $\nabla {\rho _{TE}}(x,t)$  and $\nabla {\rho _{RW}}(x,t)$ for large values of $t$ are equaled

\begin{subequations}
\label{eq:whole}
\begin{equation}
1 - \frac{{\nabla \rho _{RW}^m}}{{\nabla \rho _{TE}^m}} \approx 0.89961{\kern 1pt} \,{t^{ - 1}}, \label{subeq:1}
\end{equation}
\begin{equation}
1 - \frac{{\nabla \rho _{RW}^m}}{{\nabla \rho _G^m}} \approx 0.54298\,{t^{ - 1}}. \label{subeq:1}
\end{equation}
\end{subequations}

Let us consider now the relationship between flux and concentration gradient. 
For the Gaussian (2) the following relationship between flux and concentration gradient is valid 

\begin{equation}
\frac{1}{2}\frac{{{x_*}}}{{Dt}}\frac{{\exp \left( { - \frac{{{x^*}^2}}{{4Dt}}} \right)}}{{\sqrt {4\pi Dt} }} = \frac{\partial }{{\partial t}}\int\limits_{{x^{_*}}}^\infty  {\frac{{\exp \left( { - \frac{{{x_{}}^2}}{{4Dt}}} \right)}}{{\sqrt {4\pi Dt} }}} dx.
\end{equation}

For the functions ${\rho _{TE}}(x,t)$ and ${\rho _{RW}}(x,t)$   the fluxes in the points $x*$ ($x* \le Vt $) are calculated as 

\begin{subequations}
\label{eq:whole}
\begin{equation}
{J_{TE}}({x^*},t) = \frac{\partial }{{\partial t}}\int\limits_{{x^*}}^{Vt} {{\rho _{TE}}(x,t)dx},\label{subeq:1}
\end{equation}
\begin{equation}
J{}_{RW}({x^*},t) = \frac{\partial }{{\partial t}}\int\limits_{{x^*}}^{Vt} {{\rho _{RW}}(x,t)dx}.  \label{subeq:2}
\end{equation}
\end{subequations}

The calculations lead to the following dependencies of the square differences of the functions 	${\rho _G}(x,t)$,${\rho _{TE}}(x,t)$, ${\rho _{RW}}(x,t)$, their gradients and the corresponding fluxes at $V=1$:

\begin{subequations}
\label{eq:whole}
\begin{equation}
I_{RW - TE} = \sqrt {\int\limits_0^{x = t} {{{\left( {\rho _{RW}^{} - {\rho _{TE}}} \right)}^2}} dx} \sim {t^{ - 1.272}}, \label{subeq:1}
\end{equation}
\begin{equation}
I_{RW - G} = \sqrt {\int\limits_0^{x = t} {{{\left( {{\rho _{RW}} - \rho _G^{}} \right)}^2}} dx} \sim {t^{ - 1.255}},\label{subeq:2}
\end{equation}
\begin{equation}
I_{TE - G} = \sqrt {\int\limits_0^{x = t} {{{\left( {{\rho _{TE}} - {\rho _G}} \right)}^2}} dx} \sim {t^{ - 1.269}}, \label{subeq:3}
\end{equation}
\begin{equation}
I_{\nabla RW - \nabla TE} = \sqrt {\int\limits_0^{x = t} {{{\left( {\nabla {\rho _{RW}} - \nabla {\rho _{TE}}} \right)}^2}} dx} \sim {t^{ - 1.626}}, \label{subeq:4}
\end{equation}
\begin{equation}
I_{\nabla RW - \nabla G} = \sqrt {\int\limits_0^{x = t} {{{\left( {\nabla {\rho _{RW}} - \nabla {\rho _G}} \right)}^2}} dx} \sim {t^{ - 1.7335}}, \label{subeq:5}
\end{equation}
\begin{equation}
I_{\nabla TE - \nabla G} = \sqrt {\int\limits_0^{x = t} {{{\left( {\nabla {\rho _{TE}} - \nabla {\rho _G}} \right)}^2}} dx} \sim {t^{ - 1.626}}, \label{subeq:6}
\end{equation}
\begin{equation}
I_{{J_{RW}} - {J_{TE}}} = \sqrt {\int\limits_0^{x = t} {{{\left( {{J_{RW}} - {J_{TE}}} \right)}^2}} dx} \sim {t^{ - 1.022}}, \label{subeq:7}
\end{equation}
\begin{equation}
I_{{J_{RW}} - {J_{G}}} = \sqrt {\int\limits_0^{x = t} {{{\left( {{J_{RW}} - {J_{G}}} \right)}^2}} dx} \sim {t^{ - 1.032}}, \label{subeq:8}
\end{equation}
\begin{equation}
I_{{J_{TE}} - {J_{G}}} = \sqrt {\int\limits_0^{x = t} {{{\left( {{J_{TE}} - {J_G}} \right)}^2}} dx} \sim {t^{ - 1.029}}. \label{subeq:9}
\end{equation}
\end{subequations}
One can notice that deviations of the values considered from each other are decreased with time faster than $t^{-1}$.

\section{\label{sec:level1}Discussion}

So, for the temporal intervals $t>10^{3}\Delta t$ calculating errors in replacement of the "accurate" solution    by the functions  ,  at calculations of the desired fluxes will be less then $10^{-3}$ and for the times $t>10^{6}\Delta t$ will be less then $10^{-6}$. We should remark also that absolute values of the fluxes and their concentration functions are relatively large only for the times $x < 10^{-1}t$.  For the times $t=10^2$ the flux value is becoming less then $10^{-3}$ at $x=20$,  for $t=10^3$ – is less then $10^{-3}$ at $x=200$, for $t=10^4$ – is less then  $10^{-3}$ at $x=2000$. 

Let us consider the changings of deviations   from $\left( {{j_{RW}} + \tau {{\partial {j_{RW}}} \mathord{\left/ {\vphantom {{\partial {j_{RW}}} {\partial t}}} \right. \kern-\nulldelimiterspace} {\partial t}}} \right)$  for the function ${\rho _{RW}}(x,t)$. These deviations are defined as  

\begin{eqnarray}
{I_{\nabla {\rho _{RW}} - {J_{RW}} - \tau \partial {J_{RW}}/\partial t}} = \nonumber\\
=\sqrt {\int\limits_0^{x = t} {{{\left( {\nabla {\rho _{RW}} - ({J_{RW}} + \tau {\kern 1pt} \partial {J_{RW}}/\partial {\kern 1pt} t} \right)}^2}} dx} 
\end{eqnarray}
It was proved that function (24) is decreased quickly with growth of the time as 
${I_{\nabla {\rho _{RW}} - {J_{RW}} - \tau \partial {J_{RW}}/\partial t}} \sim {t^{ - 2}}.$

We introduce for equation (4) the corrected function $F(x,t)$ by means of relationship 

\begin{equation}
{{\vec J}_{RW}} + \tau {{\partial {{\vec J}_{RW}}} \mathord{\left/
 {\vphantom {{\partial {{\vec J}_{RW}}} {\partial t}}} \right.
 \kern-\nulldelimiterspace} {\partial t}} =  - D{\kern 1pt} F( \vec x,t)\nabla {C_{RW}}(\vec x,t)
\end{equation}
Here the index $RW$ stresses the origin of this relationship with random walk problem. Really, relationship (25) can present in the form of (4) for the random walk problem for $1D$ case with replacement of the coefficient $D$ for the parameter $D_{RW} = D{\kern 1pt} F(x,t)$ . The analysis of (25) shows that with the increasing of the observation time $F \to 1$ , but for the various values of the coordinate $x$ the times of approaching of the function $F$  to the unit value $1$are different as it is shown on Fig.3. 

\begin{figure}
\includegraphics{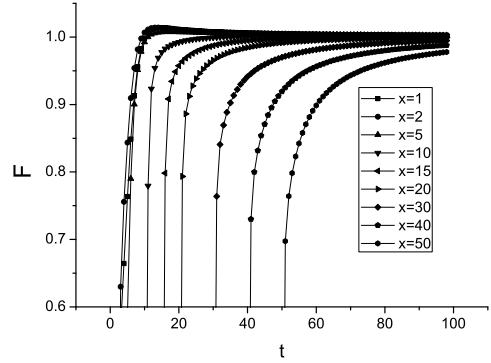}
\caption{\label{FIG3:epsart} The values of the corrected function F at various values of the current coordinate x.}
\end{figure}

We want to note here that simple and good enough analytical approximation for the function $F$ at the values $x >> 1$ and $t >> 1$ serves the following expression  

\begin{equation}
F(x,t) \approx (1 - \exp ( - {\kern 1pt} {\kern 1pt} 2Vt/\left| \vec x \right|)){\kern 1pt} \Theta {\kern 1pt} (Vt - \left| \vec x \right|)
\end{equation}
In this expression the function $F$ is presented as the approximation to the Heaviside function $\Theta$  for each fixed value of the coordinate $x$ and depends only on the characteristic velocity $V$ associated with random walk step. From this observation it is easy to conclude that for each moment of time $t_\epsilon$ if we define deviation of the function F from the unit value as $\epsilon=1-F$ the value of the corresponding coordinate  corresponding to the interval $t >> 1$ can be written as $ {x_\varepsilon } \approx \left( {{1 \mathord{\left/ {\vphantom {1 {2{\kern 1pt} V}}} \right. \kern-\nulldelimiterspace} {2{\kern 1pt} V}}} \right)\;{\kern 1pt} \left( {\;\ln {\varepsilon ^{ - 1}}} \right)\,t_\epsilon$. If we put, for example, $\epsilon = 10^{-3}$  , $V=1$ then $x_\epsilon = 0.29t_\epsilon$. It signifies that the relative velocity of approaching of $F$ to the unit value ($F \to  1$) is considerably higher in comparison with the propagation velocity of the fixed density distribution value proportional to $t^{1/2}$ .

Then equation (25) can be rewritten in the form

\begin{eqnarray}
\vec J + \tau {{\partial \vec J} \mathord{\left/
 {\vphantom {{\partial \vec J} {\partial t}}} \right.
 \kern-\nulldelimiterspace} {\partial t}} =  - D{\kern 1pt} (1 - \exp ( - {\kern 1pt} {\kern 1pt} 2Vt/\left| \vec x \right|)) \nonumber\\  \nabla C(\vec x,t)\,{\kern 1pt} \Theta {\kern 1pt} (Vt - \left| \vec x \right|)
\end{eqnarray}
where we use the notations introduced above $V= \Delta x / \Delta t$, $D=(\Delta x)^2 / (2\Delta t)$, $\tau = \Delta t /2$ , i.e. they are expressed by means of microparameters associated with random walk problem. 

We should note also the influence of fluctuations. Really, it is necessary to take into account that in derivation of equation ~(\ref{eq:1}) the locality principle is used, as well. It imposes certain limitations of the spatial scales where the diffusion type process takes place. For the equilibrium values of concentration associated with number of particles and temperature one can use evaluations of their fluctuations in correspondence with formulas

\begin{equation}
\frac{{\sqrt {\left\langle {{{\left( {\Delta N} \right)}^2}} \right\rangle } }}{N}\sim \frac{1}{{\sqrt N }},
 \frac{{\sqrt {\left\langle {{{\left( {\Delta {\text{T}}} \right)}^2}} \right\rangle } }}{T}\sim \frac{1}{{\sqrt N }}
\end{equation}

The following question can be posed, what is the size of a system should be to take into account the thermodynamic fluctuations of different macroscopic parameters as particles, temperature and etc.? It is obvious that we cannot show the accurate limits and it has a relative character. For example, for subsystem having $10^6$ particles the relative fluctuations give the value $10^{-3}$. But for small values of time and distances the thermodynamic fluctuations become very large and it is impossible to describe propagation perturbations front using only deterministic models. In other words, for the times $t < 10^3 \Delta t$  and distances $x <10^3 \Delta x$  the usage of mesoscopic approaches becomes necessary. In this case we go out from the principle locality limits and continuous medium approach. This statement coincides with the conclusions given above associated with comparison of the functions ~(\ref{eq:1}) and (5) and "true" function (14).

\section{\label{sec:level1}Conclusion}

In this paper we consider local-equilibrium and non-equilibrium models of transfer phenomena of the diffusion type. These phenomena are considered in the framework of the supposition that true solutions correspond to the solutions of the random walk problem (14) while the models (1-3) and (4-5) are considered as approximations to the "true" solution. 
The comparisons given above and analysis of the differences in fundamental solutions for equations ~(\ref{eq:1}), (5) and continuous representation of the random walk function as the density distribution function for random value (14) allow to make the following conclusions:
(a) the differences between the values of the functions considered are decreased with time as $t^{-3/2}$;  
(b) the propagation velocity of the fixed values of the functions considered is proportional to $t^{1/2}$; 
(c) on relatively large values of observation times and distances the fundamental solutions becomes very close to each other. 

Taking into account these conclusions one can formulate the criterion of applicability of the simplest equations as the telegraph equation (5) or heat transfer ~(\ref{eq:1}). Their criterion of applicability can be expressed in the form of inequalities: $t/\Delta t >>1$, $x<<t \Delta x / \Delta t$   and $t>>\tau $  where $\Delta x$, $\Delta t$, $\tau$   are defined as microparameters. We should note that for the most problems of macrophysics that conditions are satisfied for relatively small values of the observation time $t\sim 10^2 \Delta t$ . For more accurate calculations of the desired relationships between fluxes and concentration gradients in the framework of random walk problem one can use relationships (25-27). In these relationships the additional parameters figuring in diffusion equations are expressed in the form of microparameters related to the random walk problem.

\begin{acknowledgments}
We wish to acknowledge prof.R.R.Nigmatullin for useful discussions.

\end{acknowledgments}

\end{document}